\documentclass[%
 aip,
 amsmath,amssymb,
 reprint,%
]{revtex4-1}

\usepackage{graphicx}
\usepackage{dcolumn}
\usepackage{bm}

\usepackage[utf8]{inputenc}
\usepackage[T1]{fontenc}
\usepackage{mathptmx}
\usepackage{etoolbox}
\usepackage{braket}
\usepackage{hyperref}
\usepackage{natbib}
\usepackage{color}
\usepackage[bottom]{footmisc}
\usepackage[svgnames]{xcolor}
\usepackage{ulem}
\usepackage{lipsum, babel}

\makeatletter
\def\@email#1#2{%
 \endgroup
 \patchcmd{\titleblock@produce}
  {\frontmatter@RRAPformat}
  {\frontmatter@RRAPformat{\produce@RRAP{*#1\href{mailto:#2}{#2}}}\frontmatter@RRAPformat}
  {}{}
}%
\makeatother
\begin{document}

\preprint{AIP/123-QED}

\title{Public Quantum Network: The First Node}
\author{K. Kapoor\textsuperscript{*}}\affiliation{ 
Department of Physics, University of Illinois Urbana-Champaign, 1110 W Green St Loomis Laboratory, Urbana, IL, USA, 61801
}
\affiliation{ 
Illinois Quantum Information Science and Technology Center, 295 Engineering Science Building, University of Illinois Urbana-Champaign, 1101 W Springfield Ave, Urbana, IL, USA, 61801
}
\author{S.~Hoseini\textsuperscript{*}}\affiliation{ 
Department of Physics, University of Illinois Urbana-Champaign, 1110 W Green St Loomis Laboratory, Urbana, IL, USA, 61801
}
\affiliation{ 
Illinois Quantum Information Science and Technology Center, 295 Engineering Science Building, University of Illinois Urbana-Champaign, 1101 W Springfield Ave, Urbana, IL, USA, 61801
}
\author{J.~Choi\textsuperscript{*}}\affiliation{ 
Department of Physics, University of Illinois Urbana-Champaign, 1110 W Green St Loomis Laboratory, Urbana, IL, USA, 61801
}
\affiliation{ 
Illinois Quantum Information Science and Technology Center, 295 Engineering Science Building, University of Illinois Urbana-Champaign, 1101 W Springfield Ave, Urbana, IL, USA, 61801
}
\author{B.~E.~Nussbaum}\affiliation{ 
Department of Physics, University of Illinois Urbana-Champaign, 1110 W Green St Loomis Laboratory, Urbana, IL, USA, 61801
}
\affiliation{ 
Illinois Quantum Information Science and Technology Center, 295 Engineering Science Building, University of Illinois Urbana-Champaign, 1101 W Springfield Ave, Urbana, IL, USA, 61801
}
\author{Y.~Zhang}\affiliation{ 
Department of Physics, University of Illinois Urbana-Champaign, 1110 W Green St Loomis Laboratory, Urbana, IL, USA, 61801
}
\affiliation{ 
Illinois Quantum Information Science and Technology Center, 295 Engineering Science Building, University of Illinois Urbana-Champaign, 1101 W Springfield Ave, Urbana, IL, USA, 61801
}
\author{K.~Shetty}\affiliation{ 
Department of Physics, University of Illinois Urbana-Champaign, 1110 W Green St Loomis Laboratory, Urbana, IL, USA, 61801
}
\affiliation{ 
Illinois Quantum Information Science and Technology Center, 295 Engineering Science Building, University of Illinois Urbana-Champaign, 1101 W Springfield Ave, Urbana, IL, USA, 61801
}
\author{C.~Skaar}\affiliation{ 
Technology Services, University of Illinois Urbana-Champaign, 1304 W Springfield Ave, Urbana, IL, USA, 61801
}
\author{M.~Ward}\affiliation{ 
Technology Services, University of Illinois Urbana-Champaign, 1304 W Springfield Ave, Urbana, IL, USA, 61801
}
\author{L.~Wilson}\affiliation{ 
The Urbana Free Library, 210 West Green Street, Urbana, IL, USA, 61801
}
\author{K.~Shinbrough}\affiliation{ 
Department of Physics, University of Illinois Urbana-Champaign, 1110 W Green St Loomis Laboratory, Urbana, IL, USA, 61801
}
\affiliation{ 
Illinois Quantum Information Science and Technology Center, 295 Engineering Science Building, University of Illinois Urbana-Champaign, 1101 W Springfield Ave, Urbana, IL, USA, 61801
}
\author{E.~Edwards}\affiliation{ 
Illinois Quantum Information Science and Technology Center, 295 Engineering Science Building, University of Illinois Urbana-Champaign, 1101 W Springfield Ave, Urbana, IL, USA, 61801
}
\author{R.~Wiltfong}\affiliation{ 
Department of Physics, University of Illinois Urbana-Champaign, 1110 W Green St Loomis Laboratory, Urbana, IL, USA, 61801
}
\author{C.~P.~Lualdi}\affiliation{ 
Department of Physics, University of Illinois Urbana-Champaign, 1110 W Green St Loomis Laboratory, Urbana, IL, USA, 61801
}
\affiliation{ 
Illinois Quantum Information Science and Technology Center, 295 Engineering Science Building, University of Illinois Urbana-Champaign, 1101 W Springfield Ave, Urbana, IL, USA, 61801
}
\author{Offir~Cohen}\affiliation{ 
Department of Physics, University of Illinois Urbana-Champaign, 1110 W Green St Loomis Laboratory, Urbana, IL, USA, 61801
}
\affiliation{ 
Illinois Quantum Information Science and Technology Center, 295 Engineering Science Building, University of Illinois Urbana-Champaign, 1101 W Springfield Ave, Urbana, IL, USA, 61801
}
\author{P.~G.~Kwiat}\affiliation{ 
Department of Physics, University of Illinois Urbana-Champaign, 1110 W Green St Loomis Laboratory, Urbana, IL, USA, 61801
}
\affiliation{ 
Illinois Quantum Information Science and Technology Center, 295 Engineering Science Building, University of Illinois Urbana-Champaign, 1101 W Springfield Ave, Urbana, IL, USA, 61801
}
\author{V.~O.~Lorenz}\affiliation{ 
Department of Physics, University of Illinois Urbana-Champaign, 1110 W Green St Loomis Laboratory, Urbana, IL, USA, 61801
}
\affiliation{ 
Illinois Quantum Information Science and Technology Center, 295 Engineering Science Building, University of Illinois Urbana-Champaign, 1101 W Springfield Ave, Urbana, IL, USA, 61801
}

\date{\today}

\begin{abstract}

We present a quantum network that distributes entangled photons between the University of Illinois Urbana-Champaign and a public library in Urbana. The network allows members of the public to perform measurements on the photons. We describe its design and implementation and outreach based on the network. Over 400 instances of public interaction have been logged with the system since it was launched in November 2023.
\end{abstract}

\maketitle

Expanding access to technologies can enable leaps in innovation and substantial broadening of applications, as demonstrated by the evolution of the cell phone.
Quantum technology based on superposition and entanglement could similarly benefit from individuals being able to `play' with the technology.
There have been public applications of quantum networks such as the Big Bell Test\cite{big_bell_test}, quantum secure voting in Chicago\cite{quantum_voting}, a quantum secure smartphone\cite{quantum_phone}, and utilizing public fibers with industry partners at Oak Ridge National Laboratory to create a commercial quantum network\cite{ORNL_EBP}. Most current quantum networks have been developed largely without a public-facing aspect \cite{network_overview}. 
We are building a quantum network in which the public can `play' with network hardware settings in order to engage the public in exploring the potential of the technology and discovering new use cases. The network also allows us to perform quantum network research in real-world environments.

The publicly accessible quantum network we are building, which we call the Public Quantum Network (PQN), connects our laboratory at the University of Illinois Urbana-Champaign (UIUC) to a publicly accessible node at The Urbana Free Library (TUFL) in downtown Urbana, IL.
At the library, members of the public explore an interactive exhibit which provides a brief history of the development of quantum technology since the early 20th century and introduces superposition, entanglement, and measurement on the network.
The exhibit culminates in a Clauser-Horne-Shimony-Holt (CHSH) inequality experiment using pairs of entangled photons created in our lab at UIUC and distributed to the library. Library visitors can choose which polarization bases to measure the photons in, allowing them to verify the existence of entanglement for themselves\cite{clauser1969proposed}. Whereas these quantum concepts are typically first discussed in senior undergraduate physics classes, the PQN aims to make these topics accessible in a hands-on way to anyone interested in exploring them.
TUFL visitors have interacted with the CHSH measurement station over 400 times since the PQN launch in November 2023.
With the TUFL node as a model, we hope to extend the PQN by implementing future nodes in libraries, museums, and schools, thereby providing opportunities for the public to interact hands-on with quantum technology and participate in its development.


\begin{figure*}
    \includegraphics[width=\linewidth]
    {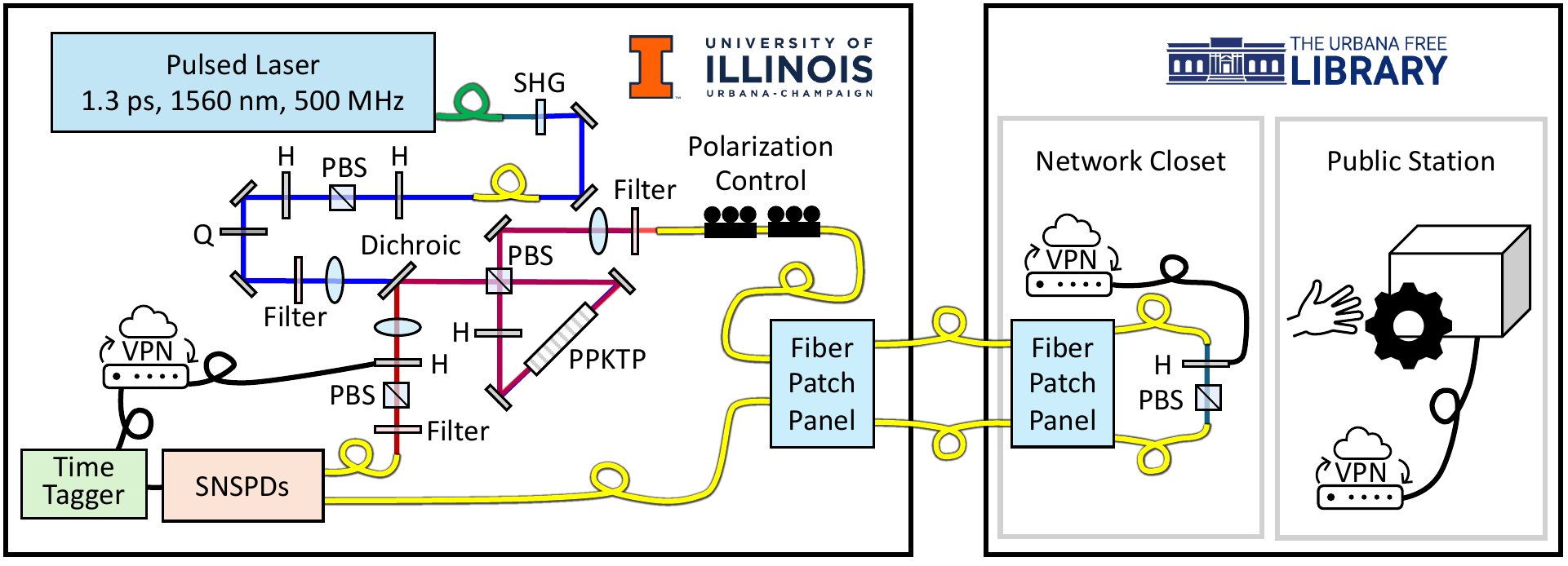}
    \caption{The experimental setup for the quantum network. A PPKTP crystal in a Sagnac interferometer is pumped in both directions to create polarization-entangled photon pairs at 1560 nm; this wavelength is compatible with deployed optical fiber. Polarization optics and a tiltable birefringent plate enable phase control before the interferometer. The idler arm is projected using a half-waveplate (HWP) and polarizing beamsplitter (PBS) before being collected into fiber and detected on superconducting nanowire single-photon detectors (SNSPDs). The signal arm is sent to the public library via a dark fiber link, launched into free space to be projected based on users' input, and sent back through fiber to the lab to be detected on the SNSPDs. The coincidences are analyzed using a time-tagger.}
    \label{fig:network}
\end{figure*}


The PQN currently comprises two network nodes, as shown in Fig.~\ref{fig:network}: one at the Loomis Laboratory of Physics (Loomis) on the UIUC campus and the other at TUFL.
These nodes are connected by two strands of dark (i.e. unused) optical fiber, forming a loop with a 12-dB loss over the 24-km round-trip distance. The optical fiber link is provided by Urbana-Champaign Big Broadband, a not-for-profit organization run in collaboration by the University of Illinois and the cities of Urbana and Champaign through the company i3 Broadband. Entangled photon pairs from a source at Loomis are distributed such that one photon from each pair remains at Loomis and the other travels through fiber to the library and back. The photon that remains at the lab is projected using a half-waveplate and polarizing beamsplitter before being collected into fiber and detected on a superconducting nanowire single-photon detector (SNSPD). The other photon is sent to the public library via the dark fiber link, launched into free space to be projected based on users' input, and sent back through fiber to the lab
to be detected on a SNSPD. The coincidences, the simultaneous detection of photons at different detectors, are analyzed using a time-tagger.



For the entanglement source, we pump a 3-cm long periodically-poled potassium titanyl phosphate (PPKTP) crystal (Raicol) in a Sagnac loop using a 1.3-ps pulsed 1560-nm laser at a repetition rate of 500~MHz (Pritel) which is then upconverted to 780-nm via second harmonic generation (SHG) (Pritel) to produce, via type-II spontaneous parametric down-conversion (SPDC), pairs of polarization-entangled photons at 1560~nm in the state \cite{sagnac_Kim,Predojevic:12}:
\begin{equation}
\ket{\Psi} = \frac{1}{\sqrt{2}}\left ( \ket{HV} + \ket{VH} \right ) = \frac{1}{\sqrt{2}}\left ( \ket{DD} - \ket{AA} \right )
\label{eq:psi}
\end{equation}
with a fidelity of over 90\%.
Since the PPKTP crystal has a manufactured specified damage threshold of >15~mW at the repetition rate of our laser, we pump the clockwise and counterclockwise directions in the Sagnac loop with 7.5~mW each to maximize the photon-pair generation rate of the source. The signal photon from each entangled pair is sent through TUFL while the idler photon is retained at Loomis. We obtain coincidence count rates in the lab of around 3,000 counts per second, with a heralding efficiency of $\sim$5\%, likely limited by imperfect spatial overlap of photons at the output port of the Sagnac loop.


\begin{figure}
    \centering
    \includegraphics[width=\linewidth]{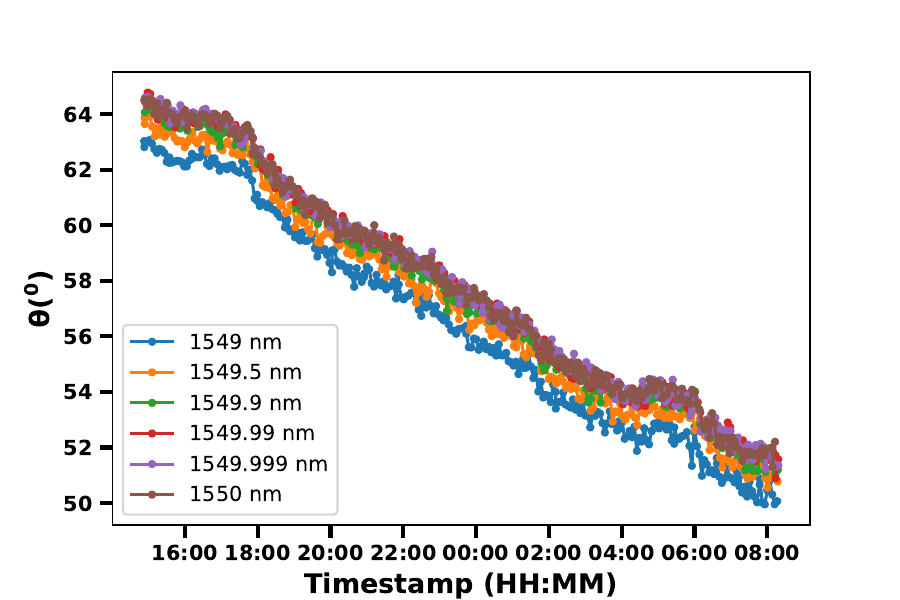}
    \caption{Drift of the azimuthal angle of the polarization, $\theta$, over the course of 18 hours on the whole 24-km fiber loop between Loomis and TUFL for different wavelengths across the bandwidth of our photon-pair source.}
    \label{fig:pd}
\end{figure}

In addition to optical losses in the deployed fiber, thermal and physical stresses induce a wavelength-dependent time-varying unitary transformation on the polarization state of photons traveling between Loomis and TUFL.
We compensate for this polarization drift by applying an inverting unitary transformation, set by adjusting manual fiber polarization controllers to minimize $\ket{HH}$ and $\ket{DA}$ coincidence counts in order to maximize the entanglement visibility (fringe contrast in the superposition basis) of the source\cite{pol_drift}.
After an initial manual adjustment of the polarization controllers, we measure a drift of $<2$ degrees per hour, as shown in Fig.~\ref{fig:pd}.

\begin{figure}
    \centering
    \includegraphics[width=\linewidth]{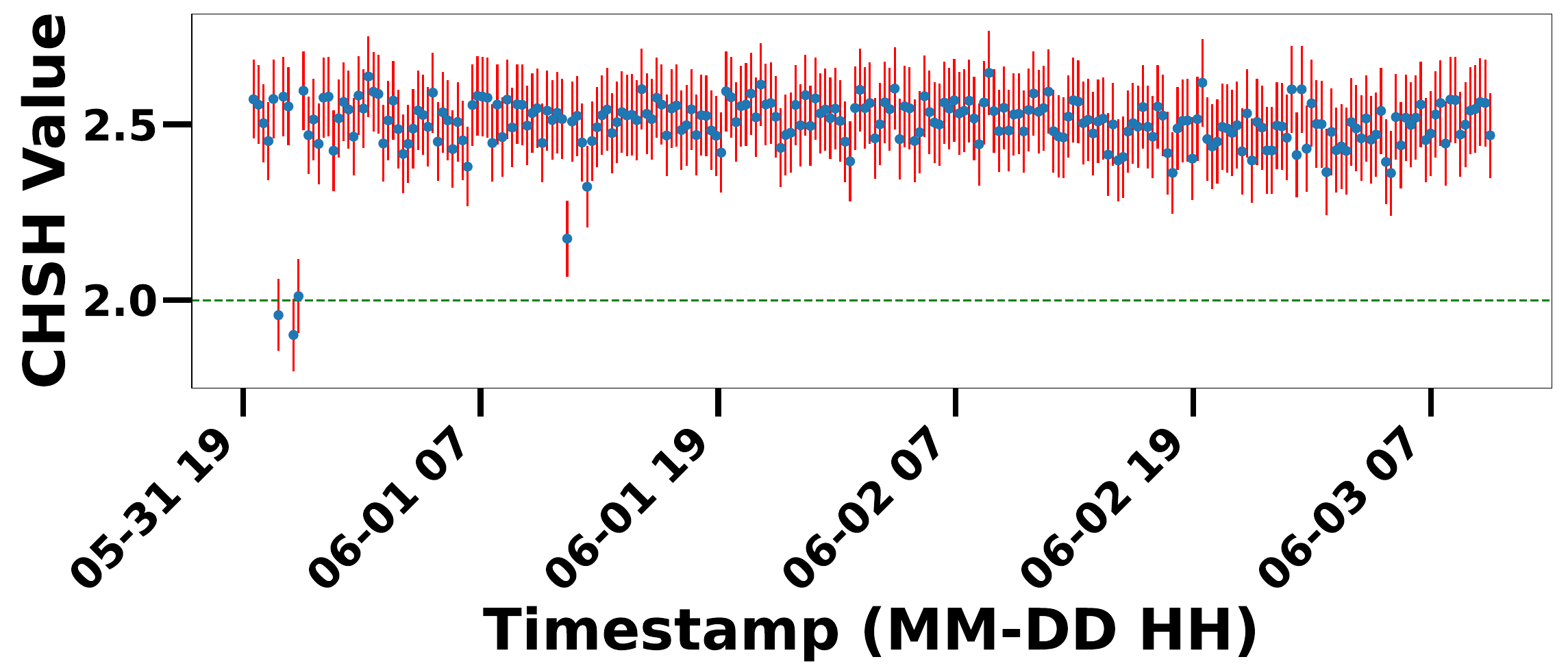}
    \caption{CHSH value over the fiber link, measured over the course of 2 days. The green dashed line marks the local realism bound of 2, the blue dots are the measured CHSH values, and the red lines are the measurement error of the CHSH values. The measured CHSH values remain above 2 for two days without active optimization, with a mean value of 2.5 with a standard deviation of 0.08. The points below 2 are believed to be due to some short-term polarization effects such as a person opening/closing a door in the network closet at TUFL.}
    \label{fig:24_hr}
\end{figure}

Both the Loomis and TUFL nodes are capable of projective polarization measurements using a half-waveplate and polarizing beamsplitter. Users at TUFL may choose any linear polarization for projection by rotating a half-waveplate in a publicly accessible setup. The angles  of linear polarization they choose are sent via a virtual private network (VPN) to hardware in the library's network closet, where the photons are launched into free space and through a motorized half-waveplate and polarizing beamsplitter before being routed back to the Loomis lab. To perform a CHSH measurement \cite{clauser1969proposed}, users are guided in written instructions to submit their choice of two angles onto which to project the photons that travel through the library, corresponding to angles $a$ and $a\;'$ in the CHSH equation, \begin{equation}
|S| = |E(a,b)-E(a,b\;')+E(a\;',b)+E(a\;',b\;')| \leq 2.
\label{eq:chsh}
\end{equation}
These angles are shared using the VPN with the lab at Loomis, where the Loomis projection angles are offset from the library angles by $22.5^\circ$ (e.g.~if the user in the libary chooses angles $a=0^\circ$ and $a\;'=10^\circ$, then the Loomis angles will be $b = 22.5^\circ$ and $b\;' = 32.5^\circ$).

The VPN link allows control over the classical hardware needed to perform experiments. A diagram of the VPN connections is provided in Supplementary Materials.
When running a particular experiment, messages are first sent to the relevant nodes, which rotate their waveplates to the required angles corresponding to the desired bases for the first measurement.
Once the waveplates are in position, another message triggers the time tagger (IDQuantique IDQ900) to start counting events from the superconducting nanowire single-photon detectors (Quantum Opus).
The devices continue to exchange messages over the VPN, alternating between setting waveplate angles and counting photons, until all 16 required measurements have been collected for the CHSH experiment.
After an initial adjustment of the manual polarization controllers to correct for the polarization transformation in the fiber by maximizing the entanglement visibility, we can observe a stable violation over the local realism bound of 2 for over two days, as seen in Fig.~\ref{fig:24_hr}.



To inaugurate the creation of this publicly accessible quantum network, a launch event was hosted at The Urbana Free Library on November 4, 2023. 
A presentation was given on the quantum network, covering relevant concepts such as polarization, superposition, and entanglement, followed by a live demonstration of the quantum link through a CHSH inequality experiment.
American Sign Language interpretation was provided during the presentation to support accessibility.
Before and after the presentation, attendees could visit eight science fair-style tables with posters and demonstrations mainly covering relevant optical and quantum technology concepts such as photons, polarization,  superposition, entanglement, photon interference, optical fibers, philosophical implications of quantum mechanics, the future of quantum networks, the importance of education and access for quantum technology, and a swag table. 
Several activities  appropriate for all ages were provided, including quantum-themed games, liquid nitrogen ice cream, and a custom-designed  coloring book on quantum networks\footnote{\url{https://iquist.illinois.edu/pqn/coloring-book}}.

For the launch event, we moved the fiber loop out from the TUFL network closet into public space. During the presentation we displayed the live count rate from the time tagger back at Loomis, and blocked and unblocked the free-space gap in the loop to change the count rate in real time, thereby demonstrating for the audience that there really were single photons travelling through the fiber.
With a confirmed optical link between Loomis and TUFL, we proceeded with the first use of the PQN by a member of the public: Leon Wilson, head of information technology at TUFL, physically rotated a half-waveplate to choose two polarization angles at which to measure the TUFL photons. The waveplate settings were relayed to a motorized rotation mount holding a half-waveplate in the free-space path of photons travelling through the library. The settings were used to demonstrate entanglement via a CHSH inequality experiment.
Following a brief delay to perform the required measurements, the experiment yielded a value of $2.39\pm0.06$, well above the threshold of 2.00, demonstrating the presence of entanglement between Loomis and TUFL, which was enthusiastically celebrated by the attendees. We estimate the launch event drew a crowd of well over 200 visitors, with standing-room only during the presentation.
This level of attendance and degree of engagement indicates a strong public interest in quantum technology.

\begin{figure}
    \centering
    \includegraphics[width=\linewidth,trim={2cm 6.5cm 2cm 0},clip]{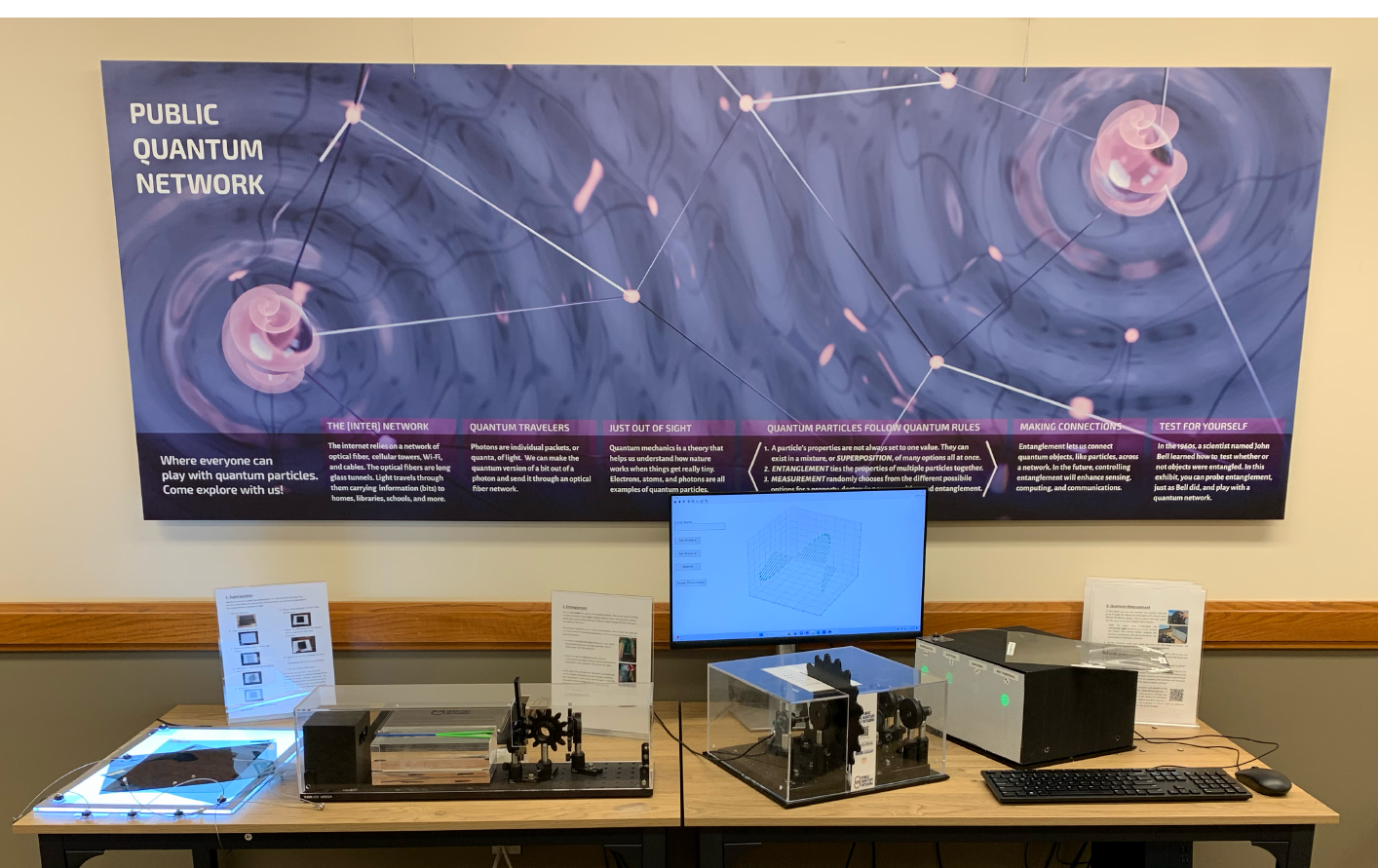}
    \caption{
        The interactive Public Quantum Network exhibit at the Urbana Free Library.
        A wall display discusses the history of quantum technology.
        Three stations engage visitors in activities about superposition (left), entanglement (left of center), and measurements on the network to perform the Bell test (right).
    }
    \label{fig:TUFL}
\end{figure}

\begin{figure}
    \centering
    \includegraphics[width=\linewidth]{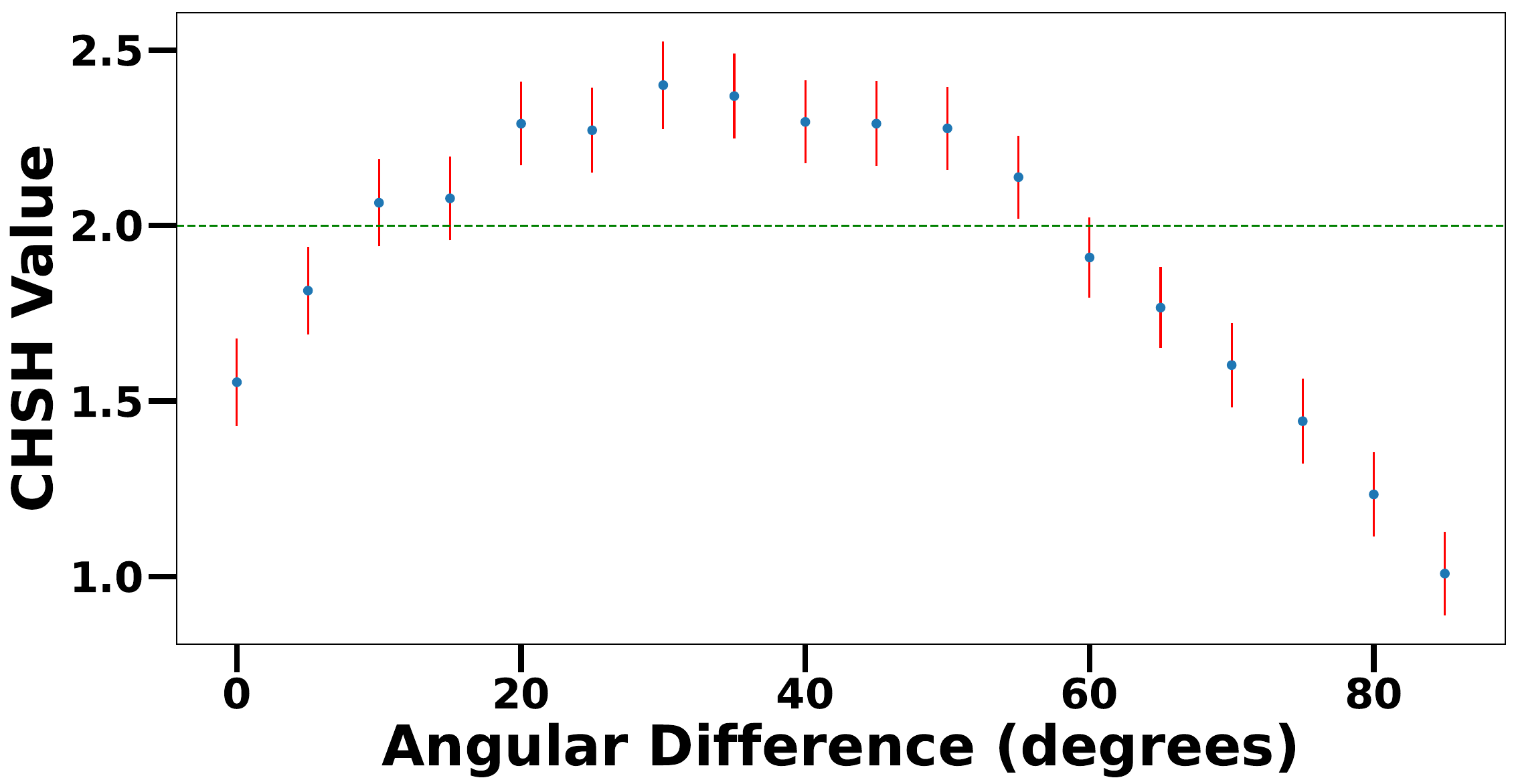}
    \caption{
        A plot of the CHSH value versus angular difference (the difference between the users' two basis choices), taken over the network after polarization correction is performed using the fiber polarization controllers. The green dashed line indicates the value 2, marking the local realism bound, and the red lines indicate measurement error.
    }
    \label{fig:CHSH_ang_dep}
\end{figure}

Beyond the displays and attractions set up ad-hoc for the launch event,
we have installed an interactive pedagogical exhibit at TUFL, shown in Fig.~\ref{fig:TUFL}, which guides visitors through three stations to learn about superposition, entanglement, and making measurements on the network.
Text guides at each station provide instructions and relevant background on each topic.

The first station introduces the concept of superposition and comprises a light table and three sheet polarizers.
By rotating one polarizer on top of the light table, users observe that about half of the light is transmitted through the sheet, regardless of angle.
The light emitted from the table is unpolarized, i.e, an incoherent mixture of multiple polarization states; performing a measurement with the polarizing sheet means that only light polarized along the direction of the sheet is transmitted.

When adding a second polarizer, users observe that the transmission through both sheets changes depending on the relative angle; in particular, they observe that \emph{no} light is transmitted through two orthogonal polarizers.
After measuring the polarization with the first sheet, the light is now polarized in the direction of that sheet, but it can also be described as a being in a superposition of the polarizations along and orthogonal to the direction of the second sheet; the component along the direction of the second sheet is transmitted.

The effect of a third polarizer laid on top of the first two initially appears trite; there is no light coming from where the two orthogonal polarizers overlap, and elsewhere it behaves the same as was observed before.
Yet, when inserting this third polarizer \emph{between} the two orthogonal ones, light is now transmitted where all three overlap spatially, where the intensity of light depends on the relative angle of the third polarizer, with the maximum occurring when it is diagonal with respect to the other two polarizers.

By interacting with this station, users learn that light can be in a \emph{superposition} of multiple different polarization states at once, and that measuring the polarization along a particular direction \emph{changes} the polarization of transmitted photons to be along that same direction. 
The quantum interpretation of this effect is that the initially horizontally polarized photons are in a superposition of diagonal and anti-diagonal polarizations. The diagonal polarizer then transmits half of the photons, leaving them diagonally polarized - a superposition of horizontal and vertical polarizations - with a 50\% chance to be transmitted through the final vertical polarizer, i.e., the measurement by the middle polarizer erases all "memory" of the original polarization state of the photon.

At the second station, visitors learn about entangled photon generation via spontaneous parametric down-conversion (SPDC).
Spontaneous parametric down-conversion is a random (``spontaneous''), non-linear (``parametric''), process which converts one high-energy photon into two lower-energy photons (``down-conversion'').
Visitors see three laser beams (representing the three photons involved in a given SPDC process) intersecting within an acrylic block indicating that SPDC only happens in specific materials, i.e., those with non-zero nonlinear optical susceptibility $\chi^{(2)}$.

The ``down-converted'' light beams exit the acrylic block and pass through two polarizers linked by a simple gear system ensuring they share the same polarization angle.
Since the light beams have the same polarization, when users rotate one of the polarizers (and therefore the other because of the gears) they observe the same intensity behavior for both light spots incident on the screen behind the polarizers.
Expanding on the angle-dependent intensity seen at the first (superposition) station, users now encounter a sort of ``entanglement'' where the intensities of the spots are \emph{correlated} with each other \footnote{Obviously, this classical analog fails to represent the other key feature of polarization-entangled photons, namely that initially each photon lacks a specific polarization and thus that there should be no angle of the analyzers for which the transmitted intensity is zero}.

The final station builds on the previous two by allowing users to perform a CHSH experiment on entangled photon pairs generated by the source .
Users can select the polarization measurement basis for the photons travelling through TUFL by interacting with a linear polarization analysis system.
This system, enclosed within a protective acrylic box, comprises a diode laser, a waveplate in a manual rotation mount, and other optical components (see Supplementary Materials) that spatially distribute the laser beam into four constituent polarization components---horizontal, vertical, diagonal, and anti-diagonal---which form four spots of varying intensity on a white plastic mesh screen. A thin 3D-printed spur gear fixed to the manual rotation mount protrudes from the acrylic box to allow users to manipulate the waveplate angle, thereby rotating the polarization state of the laser, which they observe on the mesh screen as a correlated variation in the intensity of the laser spots.
By rotating the waveplate until one spot is dark, users can easily select the corresponding orthogonal polarization state.


Behind the mesh screen, contained within another protective box, the transmitted light is focused onto four photo-resistors to measure the four linear polarization states of interest. 
By measuring the intensity of light at each location, the system computes the effective linear polarization angle $\theta$, given by
\begin{equation}\label{eq:pol}
    \theta = \operatorname{sign}{\left(P_d-P_a\right)}\frac{1}{\pi}\arccos{\left(\sqrt{P_h}\right)},
\end{equation}
where $P_x$ is a normalized value reported by the corresponding photo-resistor for each polarization $x$, with $d$, $a$, and $h$ corresponding to diagonal, anti-diagonal, and horizontal polarization, respectively.

The polarization angle is displayed on a touchscreen with a graphical interface that updates in real time as the user rotates the wheel. Two angles are entered for the CHSH measurement by rotating the wheel and pressing a "set angle" button on the touchscreen.
Once the basis angles have been selected, the user presses a button to initiate the exchange of messages between computers, which orchestrates the actual required measurements. For each measurement setting, data is accumulated for 10 seconds. Most of this time is split between waiting for waveplates to move into position and the detector integration time to obtain a sufficient number of photon counts.
With the completion of the last measurement in the sequence, the system computes and displays the final CHSH value on the screen for the user to view.
The basis choices and results are saved for future reference. 
Not all arbitrary basis choices will lead to a violation of Bell's inequality with CHSH value $>2$; indeed, users are encouraged to try various basis combinations to learn about how they relate to demonstrating entanglement. 
We essentially want users to recreate Fig.~\ref{fig:CHSH_ang_dep}, which presents CHSH values we obtained on the network as a function of the angular difference between the basis angles. By trying different angular differences users may realize they can maximize the CHSH value they obtain based off of their basis choices.

When the quantum network is inactive or inaccessible, such as during maintenance, upgrades, or active research in which the source and/or detectors are being used for another purpose, a measurement result derived from the data of Fig.~\ref{fig:CHSH_ang_dep} is displayed instead of a real-time measurement result. 
A disclaimer to this effect is included in the instructions. In the months following the network launch, live measurements by the public were performed with involvement of the research team during outreach events, as the fully automated system was still under construction. The research team continues to regularly perform tests, experiments, and maintenance on the quantum network, such as tests of network automation, polarization drift measurements, and deployment of software improvements. This work has limited the availability of the network for live, automated measurements. We anticipate the portion of live, automated measurements to increase as research activity on the network settles.

Since the initial Public Quantum Network launch event, visitors to the first publicly accessible quantum network node at TUFL have interacted over 400 times with the CHSH measurement station.
Ongoing efforts to improve this node include hardware and software upgrades as well as improvements in outreach methods and materials. 
We are exploring alternative entanglement source options for higher photon-pair generation rates and heralding efficiencies and eventual suitability for applications involving entanglement swapping \cite{Swapping:Pan}. More immediate upgrades include hardware alternatives and software improvements for faster rotation of waveplate mounts.
To gauge the impact of the Public Quantum Network on those who interact with it, we have prepared a survey, recently approved by the Institutional Review Board at UIUC. This survey will be included on the GUI in a future upgrade planned for early 2025.

Beyond the first PQN node at TUFL, we intend to establish additional nodes in community-oriented spaces such as libraries, museums, and schools.
The most active among these plans is a nascent collaboration with Fermi National Accelerator Laboratory (Fermilab) for a publicly accessible node at the Lederman Science Center on the Fermilab campus.
By expanding the Public Quantum Network to include new nodes in key locations, we hope to address and amplify the growing public interest in quantum technologies.


\section{References and Acknowledgements}

\subsection{Supplementary Materials}
Additional details on the source stability, measurement station, and code used in this experiment can be found in the supplementary materials. Data acquired for this experiment will be made available upon reasonable request.

\subsection{Author Contributions}
K. Kapoor, S. Hoseini, and J. Choi contributed equally to this work.

\subsection{Acknowledgements}
For their shared assistance, feedback, ideas, comments, and resources, the authors gratefully acknowledge the Urbana-Champaign Big Broadband network, in particular Tracy Smith and Paul Hixson; The Urbana Free Library, in particular Dawn Cassady and Lauren Chambers; and University of Illinois collaborators Andrew Conrad, Canaan Daniels, Brian DeMarco, Kim Gudeman, Angela Graham, Samantha Isaac, Spencer Johnson, Brittany Karki, Nicolas Morse, Michael O'Boyle, and Kelsey Ortiz.
This work was supported in part by NSF QLCI HQAN, Award No.~2016136.
The authors declare no conflicts of interest in this work.
\newline
Corresponding author: Keshav Kapoor, kkapoor2@illinois.edu

\bibliography{aipsamp}

\newpage

\section{Supplementary Materials}

\subsection{Source Characterization}
\begin{figure}[h]
    \includegraphics[width=\linewidth]{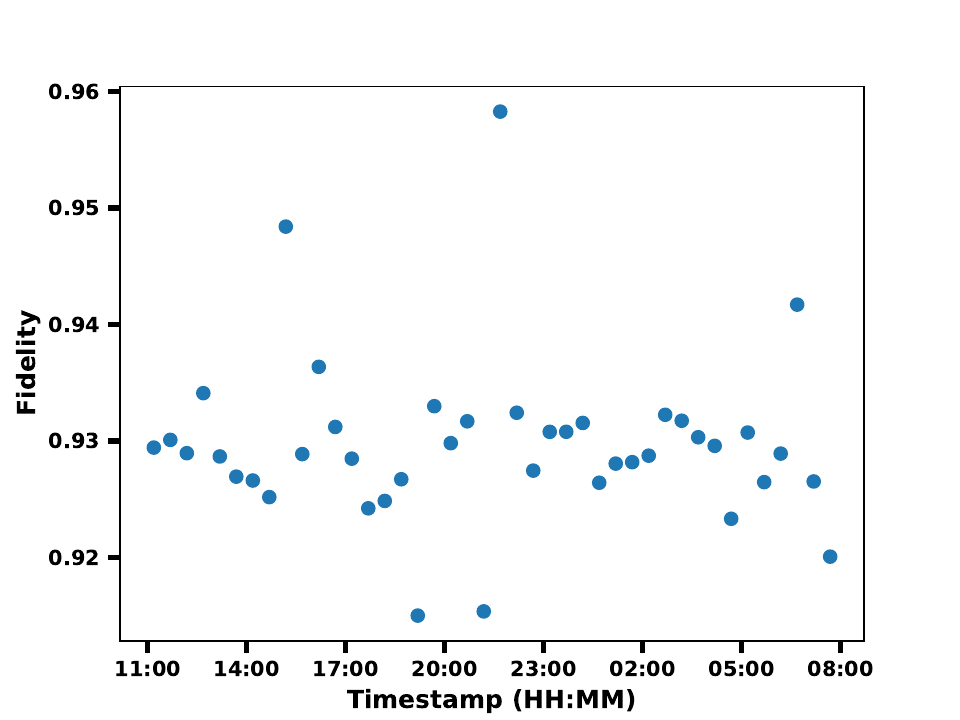}
    \caption{Entangled pair source fidelity to the target state over 20 hours.}
    \label{fig:fidelity}
\end{figure}

We characterize the fidelity of the entangled photon-pair source to the target state to determine its suitability for the Public Quantum Network.
For this measurement both photons of each pair are kept in the lab (one is not sent to the library) so as to not include effects from the fibers in the network. 
A quantum state tomography is performed on the source, once every half hour over the course of 20 hours.
From the results we calculate the fidelity with respect to the target state $\ket{\Psi} = \frac{1}{\sqrt{2}}\left ( \ket{HV} + \ket{VH} \right )$.
We obtain a fidelity of over 90\% over the course of the 20 hours, as seen in Fig.~\ref{fig:fidelity}, which indicates that our source is stable and suitable for use in our network. 

\subsection{Classical Communication Infrastructure}

\begin{figure*}[h!]
    \includegraphics[width=\textwidth]{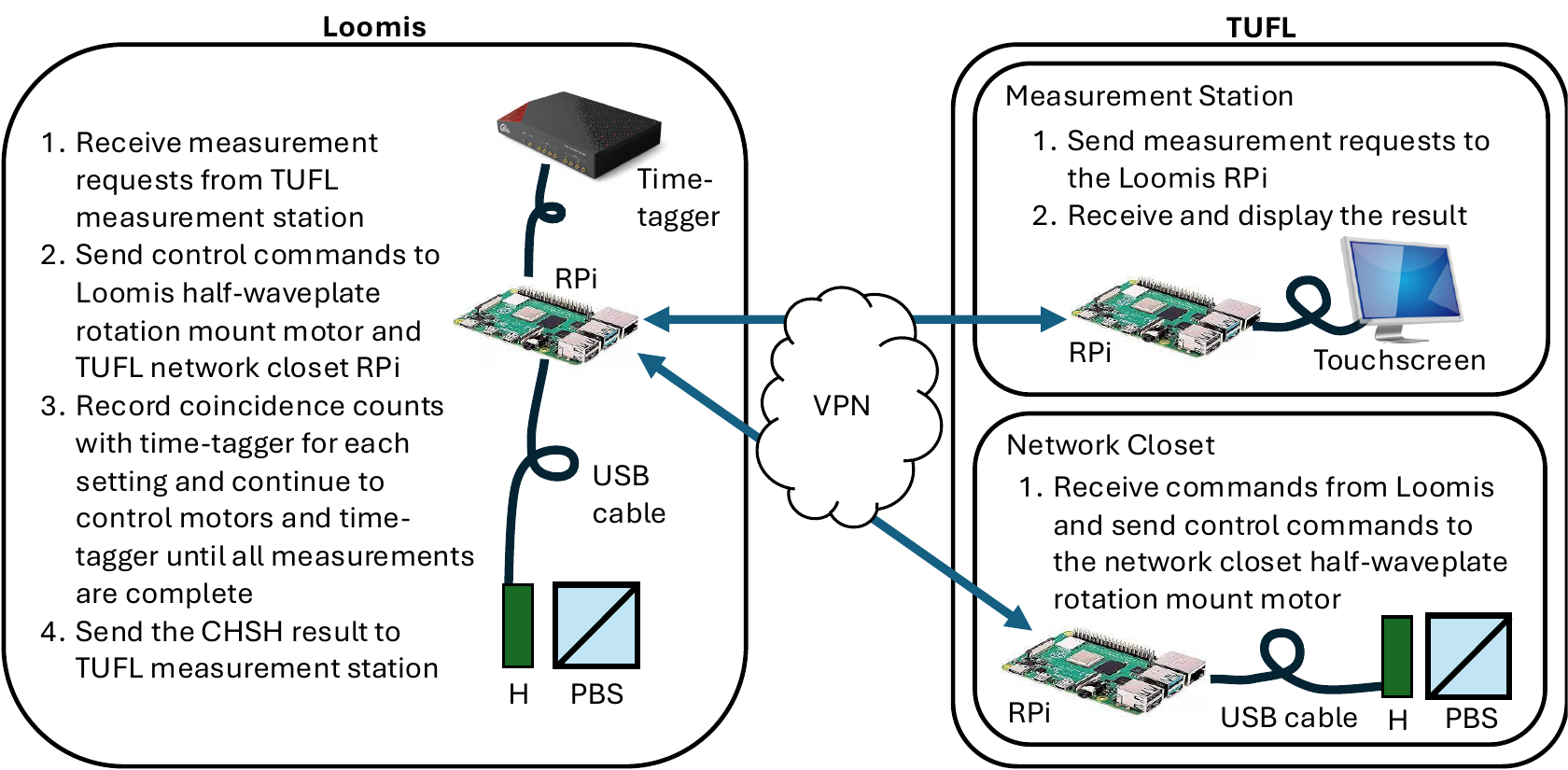}
    \caption{Public Quantum Network measurements are carried out by Raspberry Pis (RPis) in Loomis and TUFL.
    Classical messages sent over a Virtual Private Network (VPN) coordinate tasks to be executed by each RPi.}
    
    \label{fig:vpn}
\end{figure*}

\begin{figure*}
\begin{center}
    \includegraphics[width=.6\linewidth]{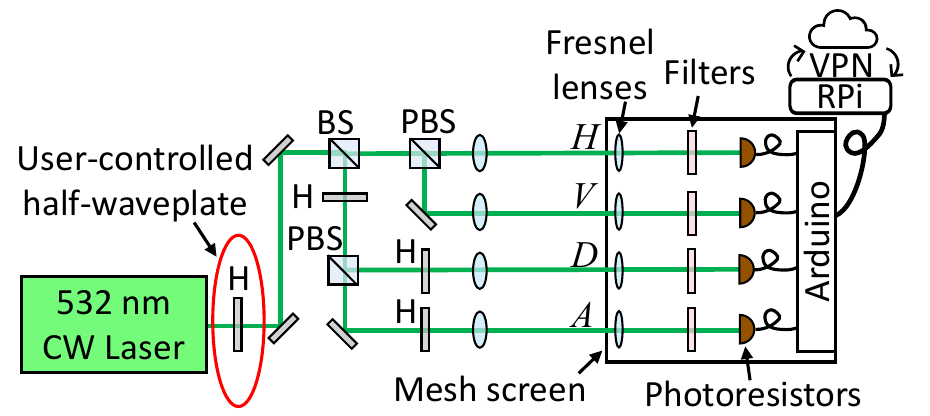}
    \caption{Schematic of the interactive measurement station at TUFL. Each path transmits specific polarizations of light: horizontal (H), vertical (V), diagonal (D), and anti-diagonal (A). The red circle indicates the half-waveplate the user can rotate. CW: continuous-wave; H: half-waveplate; BS: beamsplitter; PBS: polarizing beamsplitter; RPi: Raspberry Pi; VPN: virtual private network.}
    \end{center}
\end{figure*}

Measurements using the Public Quantum Network (PQN) are facilitated by a set of three Raspberry Pi 4 single-board computers (RPis).
One RPi at Loomis Lab provides control over waveplates to set the measurement basis and serves as an interface to the time-tagger.
At TUFL, another RPi serves a graphical user interface (GUI) application which allows members of the public to initiate a measurement using the PQN.
The third RPi controls the waveplate for measurement basis selection inside the network closet at TUFL. 
These RPis communicate with each other by sending TCP packets over a VPN (ZeroTier), as shown in Fig.~\ref{fig:vpn}.
We are in the process of developing our software into a Python package for quantum network management which will be made available on GitHub.

\subsection{Measurement Station at TUFL}
The interactive measurement station at the Urbana Free Library allows users to visualize the setting of the network closet half-waveplate that will be rotated for the quantum light. 
The green laser, after passing through the user controlled half-waveplate (controlled through the 3D printed spur gear), represents the optical path of the quantum light going into the polarization measurement optics. The optics divide the input light into four different paths corresponding to the polarization of the light.
Each path only transmits the specific polarization indicated. 
The beams in each polarization path are expanded using convex lenses to increase their visibility on a white plastic mesh screen.
The mesh screen transmits approximately 20\% of the incident light, which is subsequently focused using Fresnel lenses onto photoresistors after spectral filtering with band-pass filters.
The band-pass filters are centered at 520 nm with a full-width at half-maximum of 10 nm, ensuring that only a narrow wavelength range of the green laser is transmitted.
The intensity of the light for each polarization path is measured by the voltage changes across the photoresistors and used to reconstruct the polarization angle of the original light source, as discussed in the main text.
These voltage changes are read by an Arduino which translates the analog data into digital data read by the RPi hosting the GUI.
In order to compensate for fluctuations in the ambient light around the measurement station, a calibration procedure is performed in which the software for the measurement station tracks the highest and lowest value of each photoresistor as the user-controlled half-waveplate is rotated. The values are converted into scaling factors that are applied to subsequent readings before output.
The calibration can be redone by the user through an option on the GUI. 



\end{document}